\documentclass{article}
\usepackage{titlesec}
\usepackage[margin=1.0in]{geometry}
\usepackage{times}
\usepackage{abstract}
\usepackage{graphicx}
\usepackage{caption}
\usepackage{amsmath}
\usepackage{xcolor}

\fontsize{12}{15}
\selectfont
\date{} 
\titlespacing{\section}{0pt}{-24pt}{-\parskip} 
\titlespacing{\subsection}{24pt}{0pt}{0.5em}
\titlespacing{\subsubsection}{36pt}{0pt}{0.5em}

\titleformat{\section}{\normalfont\fontsize{12}{36}\bfseries}{\thesection.}{0.25em}{}
\titleformat{\subsection}[runin]{\normalfont\bfseries}{\thesubsection.}{3pt}{}
\titleformat{\subsubsection}[runin]{\normalfont\itshape\fontsize{12}{36}}{\thesubsection.}{3pt}{}

\title{\large\bf{Reconstruction of Interstellar Cosmic Ray Anisotropy via Heliospheric Propagation Modeling and SVD Inversion}\vspace{-12pt}}
\author{
Reilly Tom, Paolo Desiati, Juan Carlos D\'iaz-V\'elez, Perri Zilberman \\
\textit{Wisconsin IceCube Particle Astrophysics Center, Department of Physics}\\ \textit{University of Wisconsin - Madison}\\ \textit{Madison, WI, USA}\\
}

\def\bd{\begin{displaymath}}
\def\be{\begin{equation}}
\def\ed{\end{displaymath}}
\def\ee{\end{equation}}
\begin{document}
\maketitle

\begin{abstract} 
\vskip -12pt
Ground-based experiments such as the IceCube Neutrino Observatory and the High-Altitude Water Cherenkov Gamma-Ray Observatory detect cosmic-ray anisotropy on Earth, but magnetic deflections within the heliosphere complicate interpretation. Understanding the original spatial distribution of cosmic rays in the interstellar medium requires modeling particle propagation across the heliospheric boundary. We constructed a simulated transport model that backtracks antiprotons from Earth to the interstellar medium and maps spatial transfer probabilities across energies using HEALPix representations. Because our transfer matrices are singular, inversion relies on singular value decomposition regularization. Ground-based observations lose the 
equatorial North-South anisotropy
component, severely degrading interstellar map recovery. We developed a geometric circle-fitting algorithm to estimate and reintroduce the missing
component
into the observed signals. Our results show that restoring the 
North-South anisotropy
enables more accurate recovery of the interstellar dipole structure. Furthermore, Poisson statistical fluctuations on the detected event counts dominate matrix-transformation errors when spread over multiple energies.
\end{abstract}

\section{Introduction}
Earth is continuously bombarded by cosmic-ray particles across a wide energy spectrum. Cosmic rays consist of atomic nuclei that originate from powerful cosmic events, such as supernova explosions and black holes, and travel through space. Large ground-based experiments, such as the IceCube Neutrino Observatory and the High-Altitude Water Cherenkov (HAWC) Gamma-Ray Observatory, reconstruct cosmic-ray arrival directions to create all-sky distribution maps as a function of energy~\cite{Abeysekara_2019}. These measurements aim to identify cosmic-ray sources and propagation mechanics within the Interstellar Medium (ISM). However, inferring the arrival directions in the ISM is severely impeded by the heliosphere, the magnetic bubble created by the interaction between the solar wind and the ISM flow~\cite{Pogorelov_2025}.

The boundary region between the heliosphere and the ISM is likely affected by some degree of instability and turbulence.~\cite{Pogorelov_2025}. As cosmic rays enter the heliosphere, particularly at lower energy regimes (i.e., below 100 TeV), they experience significant magnetic deflections that stochastically divert their trajectories. Consequently, direct mapping of observed terrestrial signals back to ISM origins is non-trivial. To overcome this limitation, computational models must simulate particle transport across the heliospheric structure. By computing particle trajectories in a numerical model of the heliosphere~\cite{Heerikhuisen_2025}, we can quantify the heliospheric distortion and reconstruct the true cosmic-ray distribution in the local interstellar environment.

This study builds on previous research on heliospheric effects on the cosmic-ray arrival distribution on Earth~\cite{DiazVelez:201954}, using a modernized and updated heliospheric model and refining the analysis to determine the cosmic-ray directional distribution in the ISM. 
Our methodology is distinct from previous works~\cite{Zhang_2014, Zhang_2020} because it incorporates a detailed observational response function into the inversion problem. This project is the focus of a white paper for Snowmass 2021, a community-driven long-term planning initiative in particle physics organized by the Division of Particles and Fields of the American Physical Society~\cite{desiati2020snowmass}.

\section{Heliospheric Forward Simulation}
Because empirical datasets that trace individual cosmic-ray particles from the ISM to terrestrial detection do not exist, we used a computational model of the heliosphere~\cite{Heerikhuisen_2025} to numerically integrate particle trajectories simulating cosmic rays arriving at Earth. In this simulation, we back-propagated antiprotons from Earth across a broad energy range (100 GeV to 100 TeV) with a uniform directional distribution. We tracked and classified trajectories based on whether they successfully escaped to a boundary defined as the ISM. Backtracking of antiparticles is a technique used to guarantee a complete phase-space simulation connecting cosmic rays from the ISM to Earth. Assuming collisionless propagation and constant energy, these trajectories are time-reversible due to Liouville's theorem and can be used as forward-propagating protons from the ISM to Earth.

We binned trajectories reaching the ISM boundary into 10 logarithmically scaled energy bins, flanked by 2 boundary bins representing the simulation's energy bounds of 100 GeV and 100 TeV. For each energy bin, we constructed a counts mapping matrix $M$, where rows correspond to HEALPix Mollweide projection Earth pixel indices and columns correspond to ISM pixel indices.

To convert raw trajectory counts into probabilistic mapping matrices for predicting Earth anisotropy from ISM signals, each of the rows in the matrix $M$ was normalized:
\be
P_{ij} = \frac{M_{ij}}{\sum_k M_{ik}}
\ee
where $P_{ij}$ represents the normalized conditional probability that a cosmic ray arriving at Earth pixel $i$ originated from ISM pixel $j$.

\begin{figure}[htbp]
    \centering
    \includegraphics[width=1\textwidth]{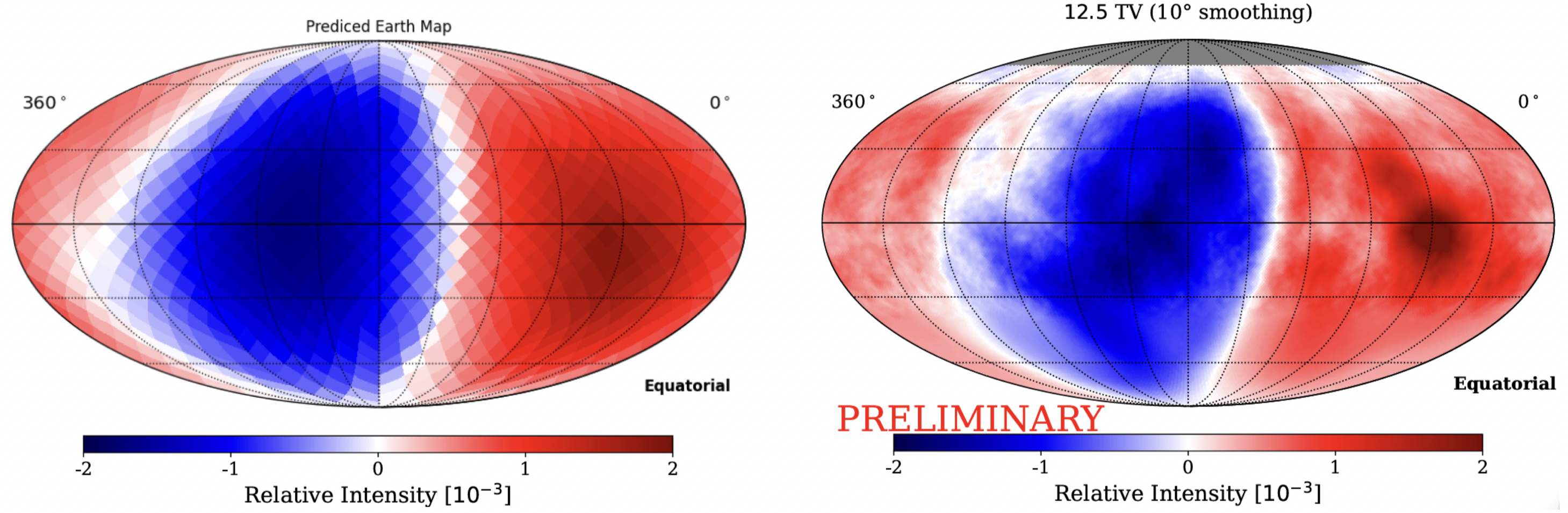}
    \caption{\textit{Left:} Predicted sky map showing the relative intensity of proton arrival distribution at Earth, derived from a dipole anisotropy in the ISM aligned with the local interstellar magnetic field. The analysis focuses on protons with a median energy of 10 TeV, following the simulated distribution observed by the IceCube Observatory~\cite{Abbasi_2025}. \textit{Right:} Measured all-sky map showing the relative intensity of the cosmic-ray arrival distribution at Earth from the combined analysis of the data collected by HAWC and IceCube at a median rigidity of 12.5 TV~\cite{DiazVelez:2025HU}.}
    \label{fig:earth_map}
\end{figure}

To incorporate the experiment's energy response function from~\cite{Abbasi_2025} over a continuous energy range, we linearly interpolated across the normalized matrices, using the logarithmic mean energy of each bin as the reference point. With this continuous operator $P(E)$, any anisotropic flux distribution in the ISM, $I_{\text{ISM}}$, can be mapped to an expected terrestrial distribution, $I_{\text{Earth}}$:
\be
I_{\text{Earth}} = P(E) \, I_{\text{ISM}.}
\ee
This forward model produces predicted Earth maps that closely match observed spatial patterns from terrestrial detectors. Figure~\ref{fig:earth_map} shows, on the left, what a 10 TeV median energy dipole anisotropy distribution in the ISM aligned with the local interstellar magnetic field, with amplitude $\delta_{\mathrm{ISM}}=2\times10^{-3}$, looks like on Earth after eliminating the observational equatorial North-South blindness, corresponding to the $m=0$ terms of the spherical harmonic functions. The figure illustrates that the simple dipole anisotropy in the ISM is distorted into a complex distribution at Earth, which resembles the experimentally measured relative-intensity sky map of the arrival distribution obtained from data collected by the HAWC and IceCube observatories at a median rigidity of 12.5 TV~\cite{DiazVelez:2025HU}, shown on the right panel of Fig.~\ref{fig:earth_map}.

\section{Inverse Reconstruction and SVD Regularization}
The primary objective is the inverse problem: recovering $I_{\text{ISM}}$ from an observed terrestrial signal $I_{\text{Earth}}$. However, the mapping matrices generated by $P(E)$ and its discrete energy counterparts are ill-conditioned. 
The complexity of the heliospheric magnetic field can lead to chaotic trajectories, even without magnetic turbulence and instabilities~\cite{Lopez-Barquero_2025}. As a result, cosmic-ray particles originating from widely separated directional pixels in the ISM may converge to the same pixel on Earth, creating degeneracy in the columns of the mapping matrix. Additionally, vast regions of the ISM have no trajectories connecting to Earth because of the characteristics of the heliospheric and uniform interstellar magnetic fields. This leads to near-zero or empty rows in the mapping matrix. 
The degeneracies in the columns of our mapping matrix, caused by these physical phenomena, mean the matrix lacks a corresponding inverse.
%

To solve this inverse problem, we employed Singular Value Decomposition (SVD) truncated regularization. To reconstruct the original cosmic-ray distribution in the ISM ($\hat{I}_{\text{ISM}}$) from our observed data ($I_{\text{Earth}}$), we solve an inverse problem governed by the forward operator $P$.
We decompose $P$ into fundamental spatial components:
\be
P = U \Sigma V^T = \sum_{k=1}^{r} u_k s_k v_k^T
\ee
where $U$ and $V$ are unitary matrices whose columns ($u_k$ and $v_k$) represent basis patterns ordered from low to high spatial frequency, $\Sigma$ is a diagonal matrix whose values ($s_k$) represent the strength (singular value) of each component, and $r$ is the matrix rank.
The reconstructed map is obtained by projecting the observed data $I_{\text{Earth}}$ onto these components and scaling by $1/s_k$:
\be
\widehat{I}_{\text{ISM}} = {P}_{K}^{\dagger}\; I_{\text{Earth}} = V_K \Sigma^{-1}_K U^T_K\; I_{\text{Earth}} = \sum_{k=1}^K \left( \frac{u_k^T\; I_{\text{Earth}}}{s_k}\right) v_k
\ee
Because small $s_k$ values (where $s_k \approx 0$) severely amplify observational noise, meaning small changes in pixel values drastically change the overall predicted map, we truncate the summation at a cutoff index $K < r$, selected via an optimization procedure. Setting $K = r$ would yield the unregularized pseudo-inverse, whereas choosing $K < r$ filters out noise to produce a stable reconstruction. We selected optimal singular value cutoffs by running Monte Carlo simulations. We optimized cutoff selection by finding the minimal Root Mean Squared (RMS) Error relative to ground-truth ISM maps across all $K=s_i$ cutoffs and validated them with a multiple hypothesis test, checking that the minimum is statistically smaller than the rest at $\alpha=0.05$ with a Holm-Bonferroni correction to control the family-wise error rate.


To incorporate simulation uncertainty into $P^{\dagger}_K$, we subjected the underlying count matrices $M$ to Poisson fluctuations before row-normalization. Inverting these fluctuated matrices incorporates transport-matrix uncertainties directly into the pseudo-inverse formulation. 

\section{Missing North-South Component and Reconstruction Algorithm}
A major challenge in observing cosmic ray anisotropy with terrestrial detectors like IceCube is the loss of the equatorial North-South component, corresponding to the $m=0$ spherical harmonic term. All ground-based observations must estimate the response function to an isotropic cosmic-ray flux, which effectively averages the cosmic-ray event counts along right ascension in equatorial coordinates. Therefore, computing the cosmic-ray intensity relative to this response function removes all anisotropic features along the equatorial North-South direction, equivalent to removing all $m=0$ spherical harmonic terms~\cite{Abeysekara_2019, Abbasi_2025, DiazVelez:2025HU}.

To evaluate the impact of this observational blindness, we ran baseline simulations using the IceCube 10 TeV median energy cosmic-ray probability mass function combined with an ISM dipole anisotropy aligned along the local interstellar magnetic field. Inverse reconstruction was performed on two signals:
\begin{enumerate}
    \item The full, uncorrupted Earth signal $I_{\text{Earth}}$.
    \item An Earth signal with the $m=0$ component explicitly removed ($I_{\text{Earth, } m\neq 0}$).
\end{enumerate}

\begin{figure}[htbp]
    \centering
    \includegraphics[width=0.8\textwidth]{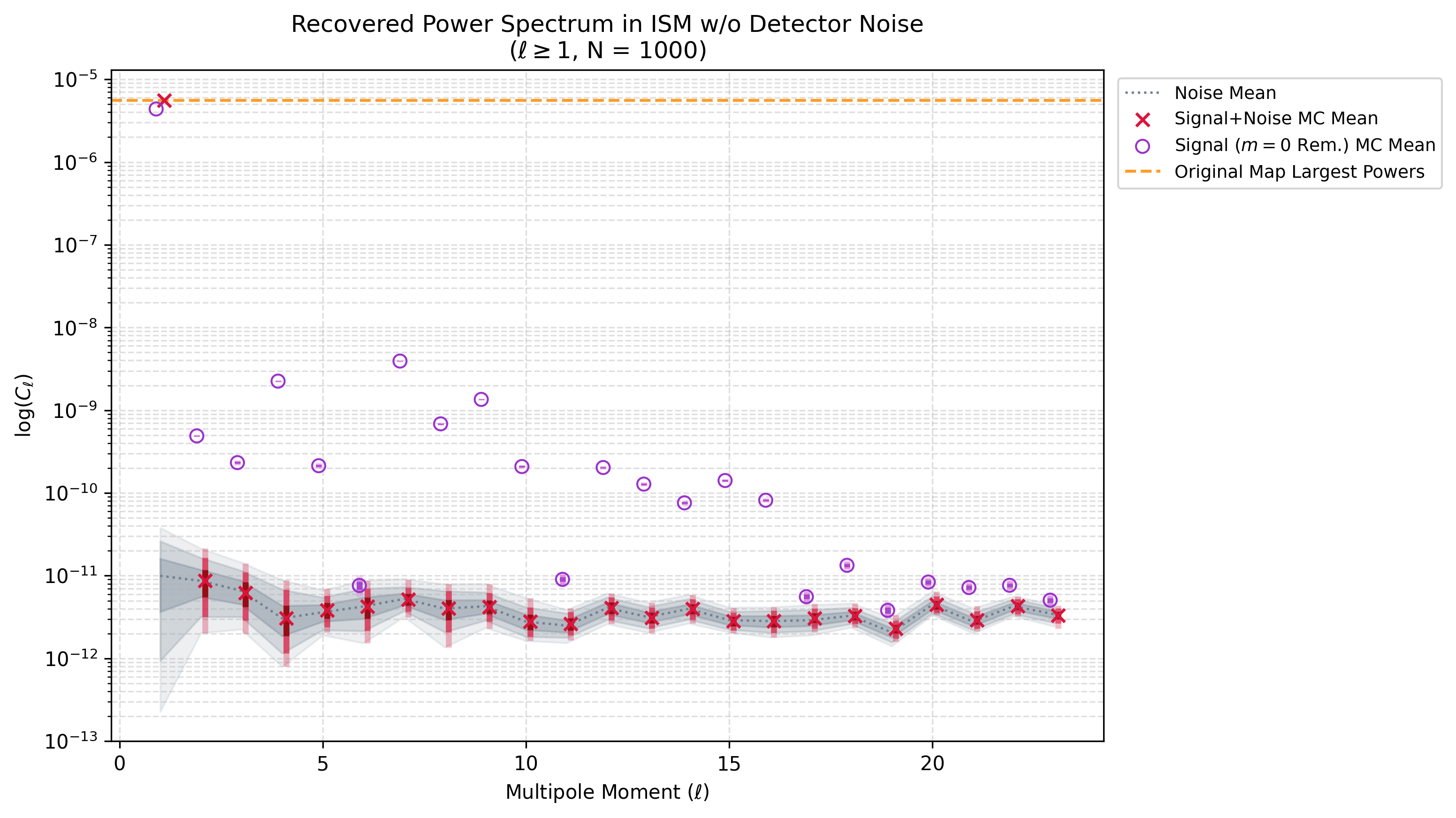}
    \caption{The Angular Power Spectrum of Monte Carlo Simulations of round-trip ISM recovery (ISM $\to$ Earth $\to$ ISM). Each data point represents the average recovered power per multipole, with the dipole power of the original ISM map being labeled as a horizontal yellow line at the top. All powers of higher-order multipole terms in the original ISM map are substantially smaller and thus off the graph.}
    \label{fig:angular_power_spectrum}
\end{figure}

Figure~\ref{fig:angular_power_spectrum} illustrates the angular power spectra ($C_\ell$) for the ISM maps, illustrating the relative significance of different angular structures in the cosmic-ray distributions. This shows that removing the $m=0$ term hinders accurate pseudo-inversion of the original ISM dipole anisotropy, severely distorting recovered multipoles. While the dipole term (multipole moment $\ell$ = 1 in the figure) can be recovered almost perfectly from a complete sky map (the red crosses in the figure) recovering the original ISM dipole (the dashed orange line), missing $m=0$ modes significantly degrade prediction fidelity (the purple circles in the figure).

\begin{figure}[htbp]
    \centering
    \includegraphics[width=0.6\textwidth]{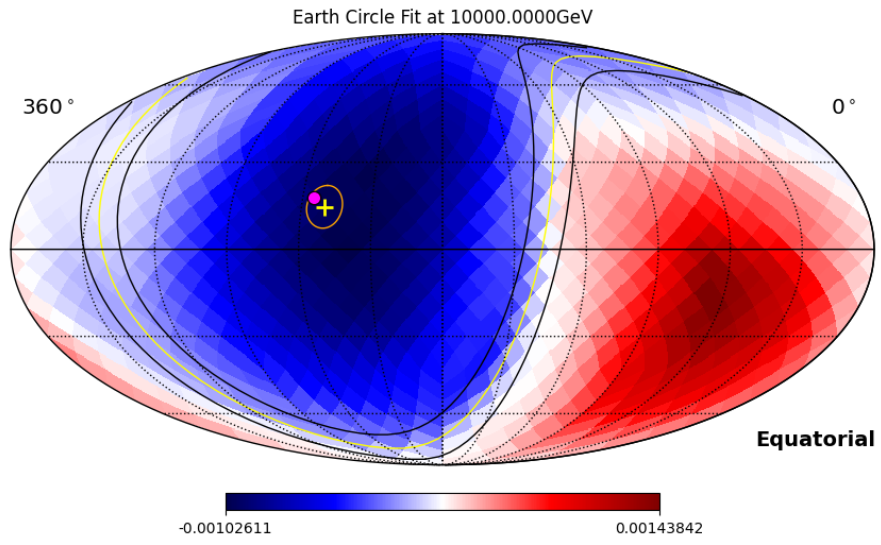}
    \caption{Demonstration of the geometric boundary circle fitting procedure on a representative sky map ($\text{NSIDE} = 8$). The fitted interface boundary and its center are shown in yellow, with surrounding black or orange bounds indicating pixel-discretization uncertainty. The purple marker denotes the true orientation of the local interstellar magnetic field vector used to validate the geometric fit. The sky map shows the result after adding the fitted $m=0$ component.}
    \label{fig:circle_fit}
\end{figure}

To address this loss, we developed a geometric reconstruction algorithm to estimate and restore the missing $m=0$ mode before matrix inversion. The algorithm follows a similar approach that was used in~\cite{Abeysekara_2019, DiazVelez:201954}, based on the assumption that the dipole anisotropy in the ISM is aligned with the local interstellar magnetic field, and operates as follows:
\begin{enumerate}
    \item We fit a circle to the lower-right boundary interface separating the deficit and excess flux regions on the terrestrial map, as shown in Fig.~\ref{fig:circle_fit}.
    \item We determine the geometric center of the boundary circle and compare it with the known orientation of the local interstellar magnetic field vector. As discussed in Ref.~\cite{DiazVelez:201954}, the boundary curve retains information regarding the dipole anisotropy of the ISM and, consequently, about the direction of the magnetic field lines. This holds true even though the observed anisotropy at Earth may be distorted by the heliospheric magnetic field.
    \item We map the angular offset between the circle center and the magnetic field direction to compute an average replacement $m=0$ term, which is iteratively added back into the observed map.
\end{enumerate}

Figure~\ref{fig:circle_fit} shows what the final output of the circle fit may look like after following the previous steps. In the diagram, the yellow plus sign marks the fitted circle's center, and the surrounding circle shows the fit error due to pixel noise. This fit is fairly accurate, as the purple dot represents the position of the ISM magnetic field direction, which is the original direction that the map was directed in before its $m=0$ terms were removed. The surrounding yellow and black lines show the fit result, with the yellow line being the exact fit and the black lines being the pixel error based on map resolution ($\text{NSIDE} = 8$ in this case).

Restoring the $m=0$ component greatly enhances the accuracy of the pseudo-inversion, bringing the primary dipole term to levels nearly indistinguishable from those derived from uncorrupted signals, as shown on the left panel of Fig.~\ref{fig:noise_comparison}. Residual errors appear as minor power in higher-order multipoles above the baseline noise floor.

\section{Detector Noise vs. Matrix Error}
We evaluated the combined propagation pipeline under two operational scenarios: a perfect detector configuration and a noisy detector configuration. In the noisy model, we converted predicted Earth flux maps into finite event counts matching the statistics of IceCube and HAWC all-sky datasets~\cite{Abeysekara_2019}, then applied Poisson fluctuations.

\begin{figure}[htbp]
    \centering
    \includegraphics[width=1\textwidth]{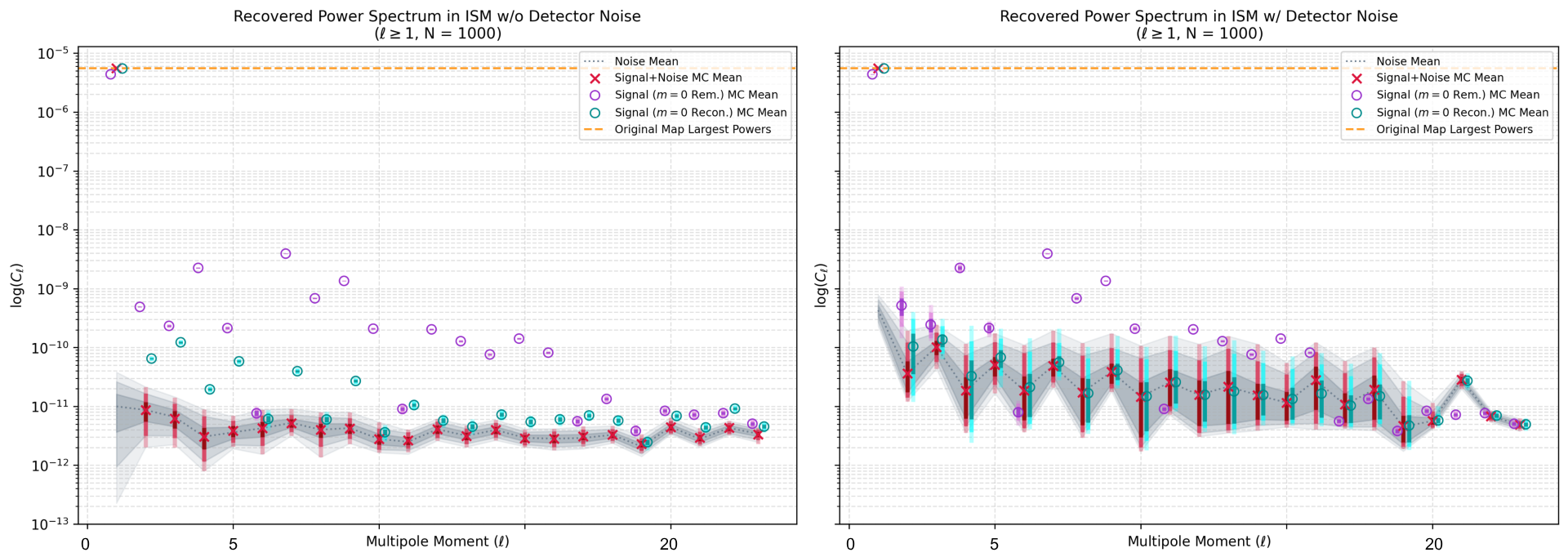}
    \caption{Comparison of reconstructed angular power spectra across Monte Carlo simulations for the round-trip propagation pipeline ($\text{ISM} \to \text{Earth} \to \text{ISM}$). \textit{Left:} Baseline recovery under an ideal detector model, accounting only for matrix pseudo-inversion error. Red crosses show the recovered angular power spectrum in the ISM in the ideal case with no losses from observational North-South blindness; purple circles show the angular power spectrum in the ISM with the North-South blindness; blue circles show the angular power spectrum after accounting for $m=0$ recovery in the circle-fit procedure. \textit{Right:} Realistic reconstruction incorporating both pseudo-inversion error and Poisson statistical fluctuations from finite detector counts. Across both panels, red markers denote the original injected signal, purple markers show the signal with the $m=0$ mode removed, cyan markers represent the reconstructed signal following $m=0$ circle-fit restoration, and the grey shaded region indicates the matrix noise floor. For multipoles $\ell > 1$, the red points align with the grey band because power at higher multipoles arises as an artifact of matrix pseudo-inversion rather than intrinsic power in the injected dipole.}
    \label{fig:noise_comparison}
\end{figure}

Figure~\ref{fig:noise_comparison} shows that when comparing angular power spectra across Monte Carlo trials, both perfect and noisy detector models successfully recover the primordial dipole mode at comparable rates. Importantly, statistical noise arising from finite detector counts dominates overall error, whereas error propagated through matrix pseudo-inversion is negligible by comparison when the energies are spread over the IceCube 10 TeV distribution.

\section{Conclusions}

We established a robust framework for modeling cosmic ray transport across the heliosphere to reconstruct the unperturbed interstellar anisotropy. By using HEALPix transfer matrices to model heliospheric transport continuously in energy, this approach enables accurate forward predictions of Earth arrival maps. Although matrix inversion via SVD regularization recovers the original ISM structures fairly accurately, it is highly sensitive to the missing $m = 0$ multipole modes; however, our circular boundary-fitting algorithm effectively restores these components, yielding a near-perfect recovery of the interstellar dipole. Ultimately, detector Poisson count noise, rather than errors introduced during matrix inversion, is the primary limiting factor in overall reconstruction accuracy.

\section{Outlook}

In this study, we applied the inversion method developed in our research to implement the round-trip procedure from the ISM to Earth and back, as outlined in this paper. Our next primary aim is to quantitatively assess the information loss that occurs during the described SVD pseudo-inversion process. We will analyze how this loss varies depending on the cutoff index applied to the singular value coefficients.

Following this analysis, we will utilize the optimized pseudo-inverse mapping operator, denoted as \(P^{\dagger}_K\), to derive the cosmic-ray distribution in the ISM. This derivation will be based on the all-sky maps produced from combined HAWC and IceCube observations, as referenced in~\cite{DiazVelez:201954}. By leveraging these observations, we aim to model cosmic-ray propagation and interactions within the ISM more accurately, enhancing our understanding of cosmic-ray physics and its implications for astrophysical processes.

\section*{Acknowledgments}
The authors acknowledge support from the Wisconsin Space Grant Consortium (WSGC) and computational resources provided by partner institutions, including the Center for High Throughput Computing (CHTC) at the University of Wisconsin{\textendash}Madison, Open Science Grid (OSG), Partnership to Advance Throughput Computing (PATh), Advanced Cyberinfrastructure Coordination Ecosystem: Services {\&} Support (ACCESS), Frontera computing project at the Texas Advanced Computing Center.

\bibliography{references}         

\end{document}